\documentclass[aps,pra,twocolumn,superscriptaddress,a4paper,showpacs]{revtex4}
\usepackage{bm}					
\usepackage{amsfonts}
\usepackage{amsmath}
\usepackage{graphicx}
\usepackage{array}
\usepackage{hyperref}
\usepackage{lipsum}
\usepackage{mathtools}

\newcommand{\diff}[1]{\ensuremath{\,{\rm d}{#1}}}

\newcommand{\bra}[1]{\ensuremath{\langle{#1}|}}
\newcommand{\ket}[1]{\ensuremath{|{#1}\rangle}}
\newcommand{\braket}[2]{\ensuremath{\langle{#1}|{#2}\rangle}}

\newcommand{\matel}[3]{\ensuremath{\bra{#1}\op{#2}\ket{#3}}}

\newcommand{\op}[1]{\ensuremath{\hat{#1}}}

\def\rme{{\rm e}}
\def\rmd{{\rm d}}
\def\rmi{{\rm i}}

\def\half{\frac{1}{2}}

\newcommand{\cep}[1]{\ensuremath{{\rm e}^{{\rm i}{#1}}}}

\makeatletter
\def\tr{\mathop{\operator@font tr}\nolimits}
\makeatother
\newcommand{\vect}[2]{\left(\begin{array}{lcl} {#1} \\ {#2} \end{array} \right)}

\newcommand{\mat}[4]{\left(\begin{array}{lclcl}{#1}&{#2}\\{#3}&{#4} \end{array} \right)}

\newcommand{\ii}{\'{\i}}

\newcommand{\be}{\begin{equation}}
\newcommand{\ee}{\end{equation}}
\newcommand{\ba}{\begin{eqnarray}}
\newcommand{\ea}{\end{eqnarray}}
\newcommand{\bfig}{\begin{figure}}
\newcommand{\efig}{\end{figure}}
\newcommand{\bcent}{\begin{center}}
\newcommand{\ecent}{\end{center}}

\def\bfalpha{{\boldsymbol \alpha}}
\def\bfbeta{{\boldsymbol \beta}}

\newcommand{\brac}[1]{\langle\,#1\,|}

\newcommand{\sym}[2]{<{#1},{#2}>}
\newcommand{\optr}[1]{\ensuremath{\hat T_{#1}}}
\newcommand{\opref}[1]{\ensuremath{\hat R_{\mathbf{#1}}}}
\def\bfalpha{{\boldsymbol \alpha}}
\def\bfbeta{{\boldsymbol \beta}}

\def\mbfx{{\mathbf x}}
\def\mbfy{{\mathbf y}}

\DeclarePairedDelimiter\floor{\lfloor}{\rfloor}

\begin{document}

\title{Phase-space representations of SIC-POVM fiducial states}

\author{Marcos Saraceno}
\affiliation{Departamento de F\'isica Te\'orica, Comisi\'on Nacional de Energ\'ia At\'omica, Buenos Aires, Argentina}
\affiliation{Escuela de Ciencia y Tecnolog\ii a, Universidad Nacional de San Martin, San Martin, Argentina}
\author{Leonardo Ermann}
\affiliation{Departamento de F\'isica Te\'orica, Comisi\'on Nacional de Energ\'ia At\'omica, Buenos Aires, Argentina}
\affiliation{CONICET, Godoy Cruz 2290 (C1425FQB) CABA, Argentina}
\author{Cecilia Cormick}
\affiliation{IFEG, CONICET and Universidad Nacional de C\'ordoba, Ciudad Universitaria, C\'ordoba, Argentina}

\date{\today}

\begin{abstract}
The problem of finding symmetric informationally complete POVMs (SIC-POVMs) has been solved numerically for all dimensions $d$ up to 67 (A.J. Scott and 
M. Grassl, {\it J. Math. Phys.} 51:042203, 2010), but a general proof of existence is still lacking.
For each dimension, it was shown that it is possible to find a SIC-POVM which is generated from a fiducial 
state upon application of the operators of the Heisenberg-Weyl group. We draw on the numerically determined fiducial 
states to study their phase-space features, as displayed by the characteristic function and the Wigner, 
Bargmann and Husimi representations, adapted to a Hilbert space of finite dimension. We analyze the phase-space localization of fiducial states, and observe that the SIC-POVM condition is equivalent to a maximal delocalization property. Finally, we explore the consequences in phase space of the conjectured 
Zauner symmetry. In particular, we construct an Hermitian operator commuting with this symmetry that leads to a representation of 
fiducial states in terms of eigenfunctions with definite semiclassical features.
\end{abstract}
%
%
\maketitle

\section{Introduction}

The question of the existence of symmetric informationally complete positive-operator valued measures (SIC-POVMs)  
\cite{Zauner_2011, Renes_2004, Appleby_2005, Scott_2006} can be
mapped to a variety of equivalent problems which have been investigated for many years eluding a conclusive answer. 
Initially posed as a mathematical problem of finding equiangular lines \cite{Lemmens_1973}, this line of
research has recently attracted renewed attention due to its relevance for quantum state tomography \cite{Caves_2002, Mendendorp_2011} 
and quantum cryptography \cite{Fuchs_2003}.

In his PhD thesis \cite{Zauner_2011}, G. Zauner 
conjectured that for any Hilbert-space dimension $d$ a solution exists which is generated by the Heisenberg-Weyl group 
acting on a {\it fiducial state},
and which has a certain additional symmetry under a unitary operator. 
The name SIC-POVM was introduced in \cite{Renes_2004}, which 
provided analytical solutions for $d$ up to 4 and
numerical solutions for $d$ up to 45. 
The case of non-prime-power dimension 6 was studied in detail in \cite{Grassl_2004},
and general covariance properties were analyzed in \cite{Appleby_2005}
and \cite{Flammia_2006}.
In \cite{Scott_2006}, the relevance of SIC-POVMs for quantum state tomography was highlighted, 
showing the optimality of this choice under statistical errors. 
The connection between SIC-POVMs and discrete Wigner functions \cite{Leonhardt_1996} was explored first in \cite{Colin_2005}, 
and later on in a generalization of the problem in \cite{Appleby_2007_OptSpec}.
The work \cite{Appleby_2007_arXiv} focused on the importance of SIC-POVMs 
from a foundational point of view concerning the structure of Hilbert space, while the relation between SIC-POVMs and mutually unbiased
bases was analyzed in \cite{Appleby_2009} for prime dimensions.
A recent computer study was reported in \cite{Scott_2010}, including numerical 
solutions up to $d=67$, together with some new analytical solutions.

The present work undertakes a new analysis of the problem from the viewpoint of phase-space representations. We consider 
the phase-space descriptions of SIC-POVM fiducial states, that is, the states that are used to generate the full POVM by application
of the Heisenberg-Weyl group operators. In particular, we explore chord, Wigner-Weyl, Husimi and Bargmann representations, 
and show how the SIC-POVM conditions manifest in phase-space. Finally, we explore the phase-space localization properties 
of fiducial states. While our study does not lead to an answer of the problem, we hope that it can provide inspiration and new insights
in the search for solutions.

This article is organized as follows: in Section \ref{sec: SIC-POVM} we give general definitions concerning SIC-POVMs and 
covariance under the Heisenberg-Weyl group. Section \ref{sec: phase space} is devoted to a review of different phase-space
representations of quantum states in a Hilbert space of finite dimension $d$, providing examples of the application of such representations to 
particular fiducial states. In Section \ref{sec: localization} we study the localization properties of 
fiducial states through their inverse participation ratio in phase space. In Section \ref{sec: Zauner} we consider a special symmetry 
conjectured by G. Zauner for SIC-POVMs in all dimensions \cite{Zauner_2011} and analyze its classical counterpart. We show that 
a Hermitian operator commuting with the Zauner symmetry can be constructed, which is a variant of the Harper Hamiltonian and provides 
a basis to expand the fiducial states. This basis has characteristic classical and semiclassical ($d\to \infty$) features. 
Finally, Section \ref{sec: conclusions} summarizes our results. In the appendices we provide additional details of our calculations. 
For simplicity, we restrict to the simpler case of odd dimensions, so that phase-space representations are not redundant \cite{Leonhardt_1996, Miquel_2002}.  


\section{Symmetric informationally complete positive-operator valued measures} \label{sec: SIC-POVM}

\subsection{Definitions}

From now on we address only scenarios with a finite Hilbert space of dimension $d$.
A generalized measurement in quantum mechanics can be described as a {\em positive-operator
valued measure} (POVM), and requires a set of positive operators 
$\hat M_j,\, j=1,\cdots m$ fulfilling $\sum_j \hat M_j = \hat{ \mathbb{I}}$. 
Such a measurement can always be cast as a projective von-Neumann
measure in a larger Hilbert space \cite{Nielsen_Chuang}.
A set of $d^2$ operators $\hat M_j$ with the additional requirement of linear independence, $\det (A ) \ne 0$ with $A_{jk} = \tr (\hat M_j\hat M_k)$,
allows one to fully reconstruct the state $\hat \rho$ from the measured values 
$p_j=\tr (\hat M_j \hat\rho)$. The operators are then said to form an {\em informationally complete POVM} (IC-POVM) \cite{Busch_1991}.

A further requirement that the operators be proportional to one-dimensional projectors with 
uniform overlap yields
\be
  \hat M_j=\frac{1}{d}\ket{\phi_j}\brac{\phi_j}
\ee
with
\be
|\braket{\phi_j}{\phi_k}|^2=\frac{d \, \delta_{jk}+1}{d+1}
\ee
and the set is then referred to as a {\em symmetric IC-POVM} (SIC-POVM) \cite{Renes_2004}.

It is not known whether it is possible, for an arbitrary finite dimension $d$,  to find 
$d^2$ states $\ket{\phi_j}$ such that the projectors $\ket{\phi_j}\brac{\phi_j}$ 
conform a SIC-POVM. The problem consists of $d^2(d^2-1)/2$ conditions on $d^2(d-1)$
complex coefficients, which makes it overdetermined so that it may not always
 have a solution. Numerical or analytical solutions have been found for all $d=2,3\cdots 67$, and it is generally 
 believed that a solution can always be found \cite{Renes_2004, Scott_2010, Zauner_2011}.

The problem has several other facets that make it interesting in other fields \cite{Renes_2004, Scott_2010}:
\begin{itemize}
 \item A SIC-POVM is a set of $d^2$ equiangular lines in complex space $\mathbb{C}^d$.
 \item A SIC-POVM is a minimal 2-design, allowing one to compute averages over 
 the Haar measure with finite sums.
 \item A SIC-POVM is a maximally equiangular tight frame.
\end{itemize}
Furthermore, SIC-POVMs are specially relevant for quantum state tomography 
due to their robustness against statistical errors \cite{Scott_2006}.

\subsection{Group-covariant SIC-POVMs and the Heisenberg-Weyl group}

Within the available tools, the simplest and most efficient way to construct 
a SIC-POVM is to impose that the operators forming it can be obtained by the 
action of a group of unitary operators acting on a fiducial state $\ket{\phi}$ according to:
\be
\hat M_j=\hat U_j\ket{\phi}\bra{\phi}\hat U^\dagger_j \,,
~~~~j=0,\cdots d^2-1 \,.
\ee
The easiest procedure, proposed in \cite{Renes_2004}, is to choose as unitaries $U_j$ the discrete phase-space displacement operators 
$\optr{\bfalpha}$ of the Heisenberg-Weyl group \cite{Weyl_1950, Schwinger_1960}. These displacement operators are defined for 
pairs of integers $\bfalpha =(\alpha_1,\alpha_2) \in{\mathbb Z}_d^2$, so that $\alpha_1$
and $\alpha_2$ are interpreted as the magnitudes of the displacement in position 
and momentum respectively. In terms of the operators $\optr{\bfalpha}$, the SIC-POVM 
condition takes the form:
\be
\label{eq: fiducial state}
|\bra{\phi}\optr{\bfalpha}\ket{\phi}|^2=
\frac{1}{d+1}~~~~~~ \forall \, \optr{\bfalpha} \ne \mathbb{I} \,.
\ee
We note that:
\be 
|\langle\phi\vert\optr{\bfalpha}\vert\phi\rangle|^2
=\tr[ \, \hat \rho \, (\optr{\bfalpha} \, \hat \rho \, \optr{\bfalpha}^\dagger) ]
\ee
with $\hat \rho = \ket{\phi}\bra{\phi}$, so that the SIC-POVM condition is a requirement on the correlation between 
$\hat \rho$ and its translated image. This correlation must be constant for all translations that are not equivalent to the identity.
Fiducial states that satisfy it have been found algebraically for some 
values of $d$ and numerically for all values $2\le d \le 67$. 
The main purpose of this work is to explore the features of these states and 
their symmetries when they are displayed in phase space. 

We now review the definitions and main properties of the 
displacement operators and also of the reflection operators that provide the Wigner-Weyl representation of quantum mechanics. 
In the discrete case, one considers the vectors of an orthonormal basis $\{\ket{n},\, n=0,1,\cdots,d-1\}$ as discretized position 
eigenstates with periodic boundary conditions, and their Fourier transforms $\{\ket{k},\, k=0,1,\cdots,d-1\}$ as discretized 
momentum states. The elements of the two bases are related by $\braket{k}{n}=\exp(-2\pi\rmi kn/d)$. The Schwinger operators 
$\hat V,~\hat U$ implement the basic position and momentum translations as:
\begin{eqnarray} 
&&\hat V\ket{n}=\ket{n+1 \, ({\rm mod}~ d)}\\
&&\hat U\ket{n}=\omega^n\ket{n}
\end{eqnarray} 
where 
\be
\omega=\exp (2\pi\rmi/d)
\ee
is a $d$-root of unity.
By definition, $\hat U$ and $\hat V$ satisfy $\hat U^d = \hat V^d =\hat{\mathbb{I}}$. Displacement operators are then defined 
for all pairs of
integers $\bfalpha =(\alpha_1,\alpha_2) \in{\mathbb Z}^2$ as
\be
\optr{\bfalpha}=\hat V^{\alpha_1} ~\hat U^{\alpha_2}
\tau^{\alpha_1\alpha_2}\,,
\label{transop}
\ee
where
\be
\tau=\rme^{\rmi \pi (d+1)/d} \,.
\ee 
For the phases we have adopted the notation of \cite{Appleby_2005} that allows one to neatly separate the even and odd $d$ cases.
Some convenient properties of this definition are $\omega=\tau^2$, $\tau^{2d}=\tau^{d^2}=1$, valid for all $d$.

 $\optr{\bfalpha}$ is the (projective) unitary representation in
the Hilbert space ${\cal H}_d$ of the group of discrete phase-space translations with coordinates $\bfalpha=(\alpha_1,\alpha_2)$. 
The operators satisfy the properties: 
\ba
&&\optr{\bfalpha}^\dagger=\optr{-\bfalpha},
\label{transpropertiesa}\\
&&\optr{\bfalpha}\optr{\bfbeta}=\tau^{\sym{\bfalpha}{\bfbeta}} \, \optr{\bfalpha+\bfbeta},
\label{transpropertiesb}
\ea
where the symplectic product $\sym{\bfalpha}{\bfbeta}$ is defined as:
\be
\sym{\bfalpha}{\bfbeta}=\alpha_2\beta_1-\alpha_1\beta_2 = - \sym{\bfbeta}{\bfalpha} \,.
\label{symform}
\ee
The additional property
\be
\optr{{\bfalpha}+d{\bfbeta}}=\epsilon^{\sym{\bfalpha}{\bfbeta}} \, \optr{\bfalpha},
\label{transpropertiesd}
\ee
with 
\be
\epsilon=\tau^d=\left\{
\begin{array}{rr}
1 & d ~ {\rm odd} \\
-1 & \quad \quad d ~ {\rm even} 
\end{array}
\right.
\label{eq: epsilon}
\ee
means that the operators are periodic in the $d\times d$ lattice for odd
$d$ and in a $2d\times 2d$ lattice for even $d$.

In the following we treat explicitly the simpler case of odd $d$. 
In this case the division by two is unambiguously given by $2^{-1}\equiv \half \, ({\rm mod}~ d) = (d+1)/2$. One can then write
$\tau=\omega^\half\equiv\omega^{(d+1)/2}$, and rewrite $\optr{\bfalpha}$ as
\be
\optr{\bfalpha}=
\sum_{j\in {\mathbb Z}_{d}} |\, j+\alpha_1/2\,\rangle \langle\, j-\alpha_1/2\,| \, \omega^{\alpha_2 j} \,.
\label{mateltrans}
\ee
where all operations inside kets and bras are performed $\mod d$.
The $d^2$ operators are linearly independent and orthonormal under
the trace product, namely, they satisfy:
\be
 \tr\big( \optr{\bfbeta}~\optr{\bfalpha}^\dagger\big)=d\,\delta(\bfbeta,\bfalpha).
\label{eq: translations orthogonality}
\ee

The closely related reflection (or phase-space point) operators are defined by a symplectic Fourier transform of  translations 
\cite{Leonhardt_1996}:
\be
\opref{\mbfx}=\frac{1}{d}\sum_{\bfalpha\in {\mathbb Z}_{d}^2}
\omega^{\sym{\mbfx}{\bfalpha}}~\optr{\bfalpha}
\label{eq:reflections}
\ee
with $\mbfx\in{\mathbb Z}_d^2$. They are easily evaluated as
\be
\opref{\mbfx}=
\sum_{j\in {\mathbb Z}_{d}} |\,x_1+ j/2\, \rangle \langle\, x_1- j/2 \,| \, \omega^{x_2 j} \,.
\label{reftrans}
\ee
and inherit from the displacements the properties
$\opref{\mbfx}=\opref{\mbfx}^\dagger$ and $\opref{\mbfx}^2=\hat{\mathbb{I}}$ so that they are both Hermitian and unitary. 
They are also linearly independent and orthogonal 
\be
\tr\big(\opref{\mbfx}\opref{\mbfy}\big)=d\,\delta(\mbfx,\mbfy) \,.
\ee
Special values of interest are
\ba
\hat R_{0,0}&=&\sum_{j\in {\mathbb Z}_{d}}\ket{j}\bra{-j}=\frac{1}{d}\sum_{\bfalpha\in {\mathbb Z}_{d}^2}\optr{\bfalpha}\,,\\
\hat T_{0,0}&=&\sum_{j\in {\mathbb Z}_{d}}\ket{j}\bra{j}=\frac{1}{d}\sum_{\mbfx\in {\mathbb Z}_{d}^2}\opref{\mbfx}\,.
\ea

\section{Phase-space representations} \label{sec: phase space}

The continuous group of Weyl displacements and the associated set of reflections have provided a way to express the quantum-mechanical 
treatment of a particle in one spatial dimension as a phase-space theory, but with the characteristic features due to the 
uncertainty principle exactly built in. Operators, observables and quantum states can be mapped under certain restrictions 
to c-number phase-space quasi-distributions and conversely functions in phase space can be quantized as operators in Hilbert space. 
In the discrete case an analogous situation arises \cite{Leonhardt_1996, Vourdas, Miquel_2002, Rivas}.
The two unitary bases described in the previous section provide a way to represent any operator in ${\cal H}_d$ in terms of 
discrete c-number arrays with the properties of  phase-space quasi-distributions.  On the other hand continuous representations 
are also possible for operators in ${\cal H}_d$ in terms of specially adapted Bargmann and Husimi functions 
\cite{Leboeuf_1990, Nonnenmacher, Bengtsson_Zyckowski}. In the following we review briefly the definitions and discuss the features 
displayed by these representations as they pertain to the structure of SIC-POVM fiducial states.

\subsection{Discrete phase-space representations} 
Using the orthogonality condition (\ref{eq: translations orthogonality}), the 
$d^2$ linearly independent translations can be used as a basis to represent any operator 
$\hat A$ as:
\be
C_A (\bfalpha)=\tr\big(\hat A ~\optr{\bfalpha}^\dagger\big)
\label{eq: chord}
\ee
where ${\bfalpha}=(\alpha_1,\alpha_2) \in{\mathbb Z}_{d}^2$.
In terms of these coefficients the operator can be reconstructed as:
\be 
\label{eq: state chord}
\hat A =\frac{1 }{d}\sum_{\bfalpha\in\mathbb
Z_d^2} C_A(\bfalpha)~\hat T_\bfalpha
\ee
The same can be done in the reflection basis
\be
W_A(\mbfx)=\tr \big(\hat A \opref{\mbfx}\big)
\label{eq: Wigner}
\ee
and the reconstruction then reads
\be
\hat A =\frac{1 }{d}\sum_{\mbfx\in\mathbb
Z_d^2}W_A(\mbfx)\opref{\mbfx} .
\ee
$C_A(\bfalpha)$ is called the Weyl (or chord) representation of the operator, while $W_A(\mbfx)$ is its Wigner (or center) 
representation. The alternative names \cite{Ozorio_1998} arise from the fact that the transformation $\mbfx_A\mapsto \mbfx_B$ is labeled by
the chord $\bfalpha=  \mbfx_B - \mbfx_A$ in the case of translations, while for reflections it is labeled by the center $(\mbfx_A + \mbfx_B)/2$. 

Standard properties of these representations are:
\be
\tr\big(\hat A\big) = C_A(0,0)= \frac{1 }{d}\sum_{\mbfx\in\mathbb
Z_d^2} W_A(\mbfx)\label{eq: chord trace}
\ee
\ba
\tr\big(\hat A \hat B\big)&=&\frac{1}{d}\sum_{\bfalpha\in\mathbb
Z_d^2}C_A(\bfalpha)C_B(-\bfalpha)\\
&=&\frac{1}{d}\sum_{\mbfx\in\mathbb
Z_d^2}W_A(\mbfx)W_B(\mbfx)
\label{eq: chord property b}
\ea
When applied to density matrices $\hat \rho$, $C_\rho(\bfalpha)$ is usually called the characteristic function while $W_\rho(\mbfx)$ 
is the Wigner function. However the standard normalization is different. For uniformity, we prefer to keep the normalization implied 
in (\ref{eq: chord trace}). With this normalization the values of the Wigner or chord function of a normalized state $\rho$ are 
bounded as $|C_\rho(\bfalpha)|\le 1$ and $|W_\rho(\mbfx)|\le 1$.

Furthermore, for a Hermitian operator $\hat A=\hat A^\dagger$, $W_A(\mbfx)$ is real and
$C_A(\bfalpha)=C_A(-\bfalpha)^\ast$, where
$- {\bfalpha}$ is understood to be taken modulo $d$. 
Other operator features are more difficult to write in terms of these representations, in particular positivity for a density 
operator or the pure-state condition $\hat \rho^2=\hat\rho$. The latter leads to the non-linear relations:
\be
C_\rho (\bfalpha)=\frac{1}{d}\sum_{\bfbeta\in\mathbb
Z_d^2}C_\rho (\bfbeta) C_\rho (\bfalpha-\bfbeta)\tau^{\sym{\bfalpha}{\bfbeta}}\,,
\label{eq:purechord}
\ee
\be
W_\rho(\mbfx)=\frac{1}{d^2}\sum_{\mbfx_1\mbfx_2\in\mathbb
Z_d^2}W_\rho(\mbfx_1)W_\rho(\mbfx_2)\omega^{2\sym{\mbfx-\mbfx_1}{\mbfx-\mbfx_2}}\,.
\label{eq:purewig}
\ee

The defining Eq. (\ref{eq: fiducial state}) for a SIC-POVM fiducial state imposes a necessary structure for its chord representation.
Thus a putative SIC-POVM fiducial state must be of the form:
\be
\hat \sigma=
\frac{1}{d}\Bigg[\hat{\mathbb{I}}+\frac{1}{\sqrt{d+1}}\sum_{\bfalpha\ne 0}
\rme^{\rmi\psi_\bfalpha} \, \optr{\bfalpha}\Bigg].
\label{chordstructure}
\ee
If the phases $\psi_\bfalpha$ are chosen to satisfy $\psi_\bfalpha=-\psi_{-\bfalpha}$, then $\hat \sigma$ is Hermitian and 
$\tr\hat\sigma^2=\tr \hat\sigma=1$. However, the resulting operator is not necessarily positive. Enforcing the nonlinear condition
(\ref{eq:purechord}) that makes $\hat \sigma$ a pure state shows the difficulty of characterizing (and finding) fiducial states.

In Fig. \ref{chordwigner} we show as an example the chord and Wigner representations of a fiducial state for the case $d=5$ 
(the state coefficients are taken from \cite{Scott_2010}, where the state is labeled ``5a''). Each square in the figure represents 
the value of $\psi_{\bfalpha}$ in the region $-(d-1)/2\le\bfalpha\le(d-1)/2$ for the chord function and the real values of the 
Wigner function $W_\rho(\mbfx)$ in $-(d-1)/2\le\mbfx\le(d-1)/2$.  The triangles connect equal  values of the distributions on 
the cycles of the Zauner map (see Sec.{\ref{sec: Zauner}) displaying the Zauner symmetry of the state. Notice that $W_\rho(\mbfx)$ 
was computed for this state in \cite{Colin_2005}.

\begin{figure}
\begin{center}
\includegraphics[width=0.43\textwidth]{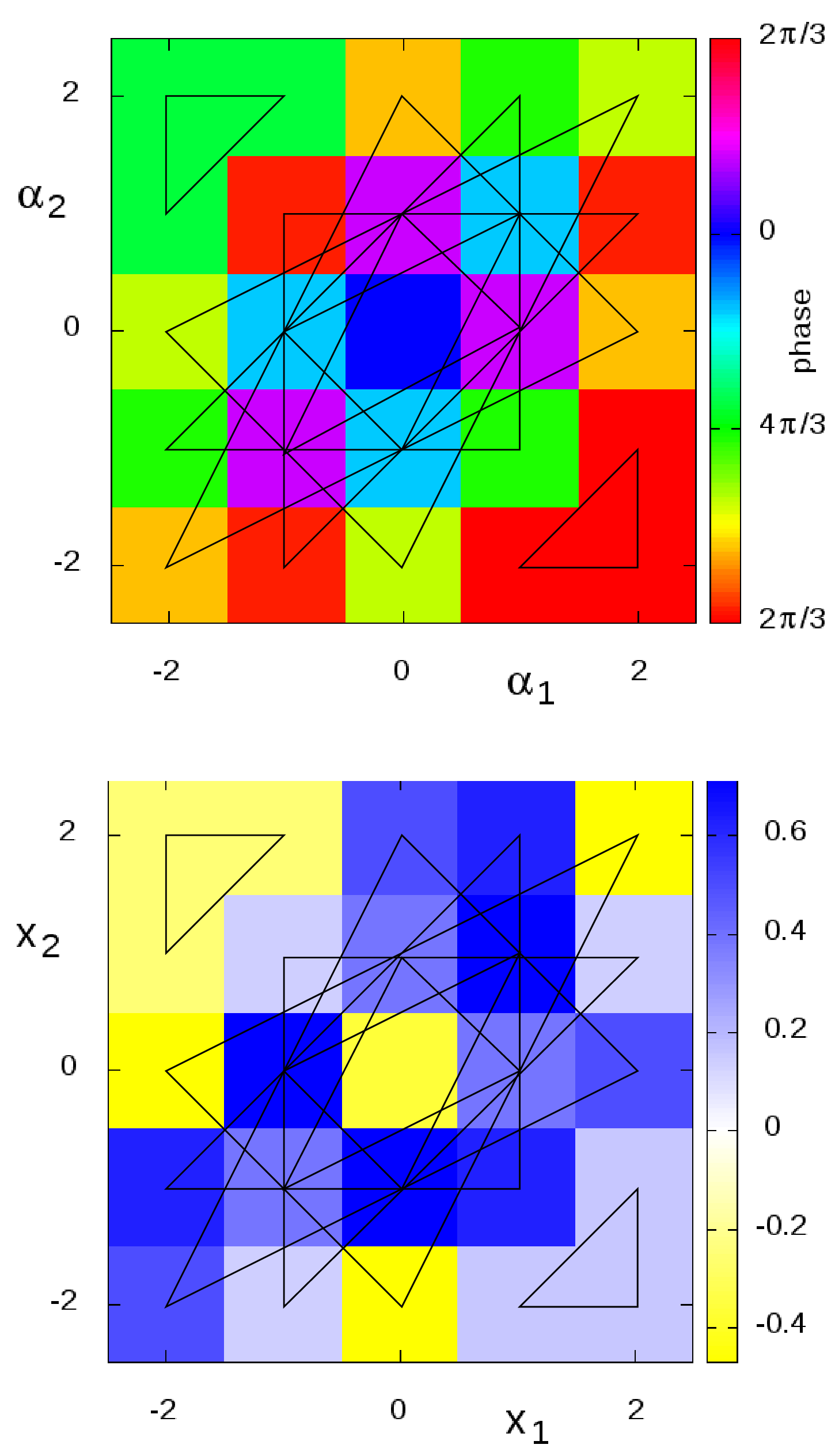}
\caption{(color online)
Chord and Wigner representations of the state ``5a'' in the list in \cite{Scott_2010} in top and bottom panels respectively. 
The color scheme adopted is  HLS, with hue and lightness labeling phase and modulus. In the bottom figure blue is positive and yellow negative.
The triangles indicate the cycles of period three of the Zauner map, connecting equal values of the distributions. (see Sec. \ref{sec: Zauner}).  
\label{chordwigner}}
\end{center}
\end{figure}

\subsection {Husimi and Bargmann representations} 

The Bargmann representation of quantum mechanics \cite{ Bargmann} maps the Hilbert space of square 
integrable functions on the real line to one of analytic (entire) functions in 
the complex plane. Namely, the Bargmann transform of a state $\ket{\psi}$, $\psi(z) = \braket{z}{\psi}$, is given by its projection over the  
coherent states $\ket{z},z\in \mathbb{C}$ with $z=q-\rmi p$, each of which is an unnormalized Gaussian wave-packet centered at the 
phase-space point $\{q,p\}$. The kernel that provides the mapping for the position kets $\ket{x}, x\in \mathbb R$ is:
\be
\braket{z}{x}=\frac{1}{(\pi\hbar)^\frac{1}{4}}
\exp\left\{\frac{1}{\hbar}\left[-\frac{(z-x)^2}{2} + \frac{z^2}{4}\right]\right\} \,.
\label{eq:coherent}
\ee
Here, $\hbar$ is left explicitly as a dimensionless measure of the elementary quantum cell in phase space.

The Husimi representation of the state is closely related with the Bargmann representation and is given by the expression:
 \be
 H_\psi(z,\bar z)\equiv H_\psi(q,p)= \frac{|\braket{z}{\psi}|^2}{\braket{z}{z}}=|\braket{z}{\psi}|^2\rme^{-|z|^2/(2\hbar)}
 \label{eq:Husimi}
\ee
where the overline stands for complex conjugation. As opposed to the Wigner function, which cannot be interpreted as a probability
distribution because it can take negative values, the Husimi representation is clearly positive by construction.

The adaptation of the Bargmann representation to ${\cal H}_d$ with a phase space with periodic boundary conditions in position and 
momentum \cite{Leboeuf_1990} (a unit torus) entails the immediate consequence that $\hbar$ can only take the values
\be
\hbar^{-1}=2\pi d
\ee
so that the unit torus is spanned by $d$ quantum states of area $2\pi\hbar$. Moreover
position and momentum eigenstates are discretized as $\ket{x}\to\ket{n/d},n\in\mathbb{Z}_d$ and
$\ket{p}\to\ket{k/d},k\in\mathbb{Z}_d$. To complete the construction the kernel is imposed to be periodic 
on the interval $0\le x \le 1$ resulting in the definition
\be
\braket{z}{n/d}_d=\sum_{\mu\in\mathbb{Z}}\braket{z}{n/d +\mu} \,.
\ee
Using (\ref{eq:coherent}) we obtain
\begin{multline}
\braket{z}{n/d}_d=\exp \left\{2\pi d \left[\frac{z^2}{4}-\frac{(z-n/d)^2}{2}\right]\right\}\\
 \theta_3 \big(\rmi \pi (n-dz)|\rmi d\big)
\label{eq:periodic}
\end{multline}
Where $\theta_3$ is the Jacobi theta function with the convention from \cite{Whittaker}:
\be
\theta_3(z|\tau)=\sum_{\mu\in\mathbb{Z}}\rme^{\rmi\pi\tau \mu^2}\rme^{2\rmi \mu z}
\ee
With this definition any state $\ket{\psi}$ in ${\cal H}_d$ with coefficients $\braket{n}{\psi}$ is mapped to the Bargmann function in the form:
\be
B_\psi(z) = \braket{z}{\psi}_d=\sum_{n\in\mathbb{Z}_d}\braket{z}{n/d}_d \braket{n}{\psi}
\label{eq:Bargmann}
\ee

According to this mapping, $\braket{z}{\psi}_d$ is an entire analytic function which moreover satisfies the following quasiperiodic 
boundary conditions in the fundamental unit cell in $\mathbb C$ \cite{Leboeuf_1990}: 
\ba
 \braket{z+1}{\psi}_d =\rme^{\pi d (\half+z)}\braket{z}{\psi}_d\,, \nonumber \\
 \braket{z+\rmi}{\psi}_d =\rme^{\pi d (\half-\rmi z)}\braket{z}{\psi}_d\,.
 \label{eq:boundary}
\ea 
Using Cauchy's theorem and these boundary conditions to integrate on the contour of the unit cell, it is easy to show that there are 
exactly $d$ zeroes in it and this pattern of zeroes is periodically repeated in the whole complex plane. The zeroes $z_i, i=1,\ldots, d$ 
in the unit cell are constrained by the relation
\be
\frac{1}{d}\sum_{i=1}^d z_i= \frac{(1+\rmi)}{2} \quad {\rm mod}(1,\rmi)
\label{eq:constraint}
\ee

\begin{figure}[t]
\begin{center}
\includegraphics[width=0.4\textwidth]{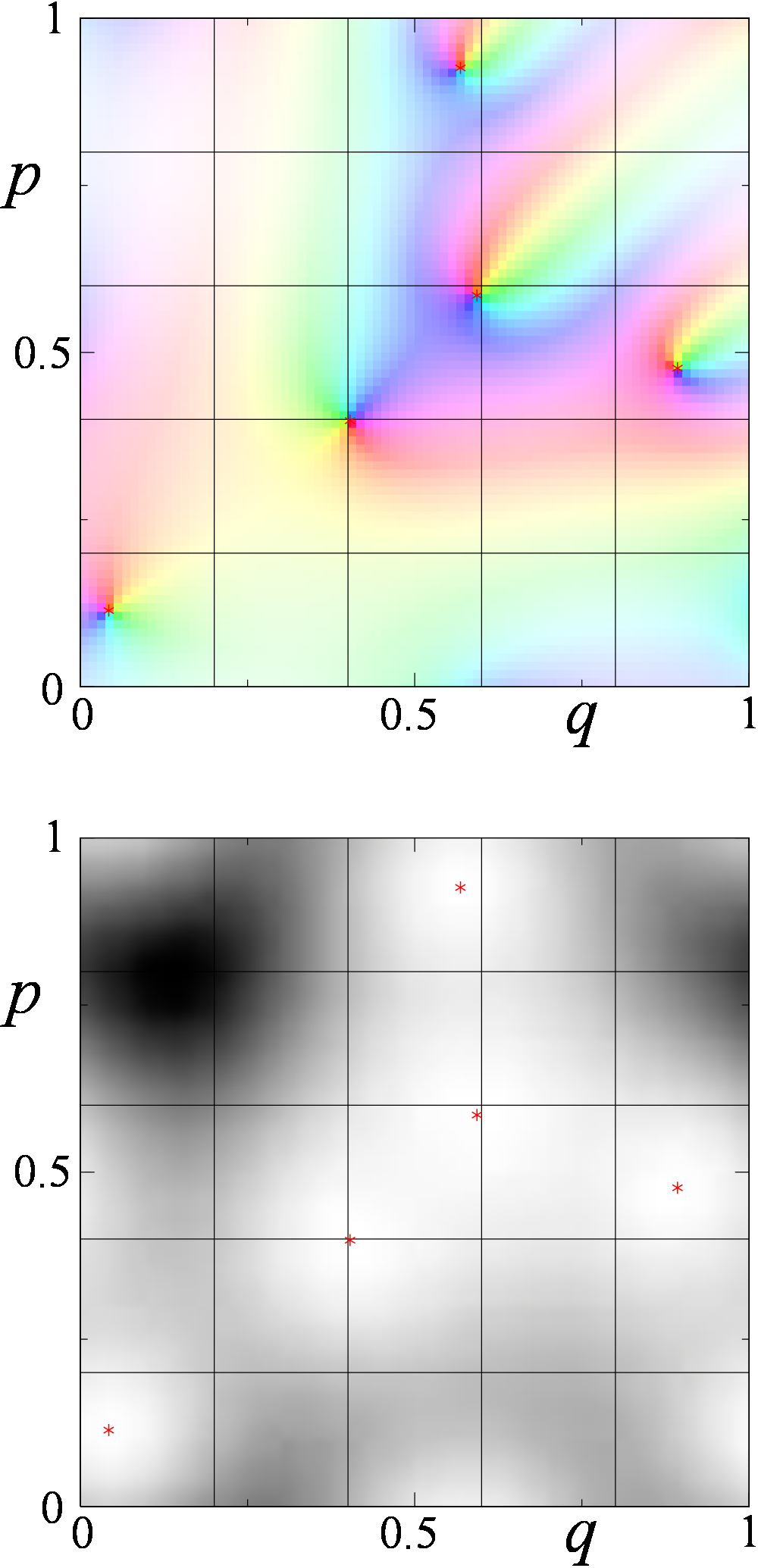}
\caption{
(color online) The Husimi and Bargmann representation of the state ``5a'' in the list in \cite{Scott_2010}. 
The bottom panel represents the Husimi density, Eq. (\ref{eq:Husimi}) with a linear intensity scale between the maximum and zero. 
The locations of the 5 zeroes are also marked. The top panel displays the phase of the Bargmann function, Eq. (\ref{eq:Bargmann}), 
in the fundamental cell. The zeroes are located at the phase dislocations. 
\label{husbarg}}
\end{center}
\end{figure}

Knowledge of the zeroes enables the full reconstruction of the quantum state via the Hadamard factorization of entire functions, 
which in our case becomes
\be
\braket{z}{\psi}_d= C(z)\prod_{i=1}^d  \braket{z+z_0 -z_i}{0}_1
\ee
Where $C(z)$ is non-vanishing and the factors are the fundamental quasi-periodic functions for $d=1$:
\be
\braket{z}{0}_1\equiv \psi_1(z)=\rme^{ -\pi\frac{z^2}{2}}\theta_3\left(-\rmi\pi z|\rmi \right)
\ee
which have a single zero at $z_0=(1+\rmi)/2$ and are peaked at $z=0$. The consequence is then that there is a one to one correspondence 
between the $ d $ complex zeroes constrained by Eq. (\ref{eq:constraint}) and the $d-1$ complex amplitudes that define the normalized
projector $\ket{\psi}\bra{\psi}$. This highly non-linear correspondence, analogous to the relationship between the coefficients and the 
zeroes of a polynomial, constitutes the $\it stellar$ representation of the pure state \cite{Leboeuf_1990, Bengtsson_Zyckowski}.

The Husimi function, now defined as
\be
 H_\psi(z,\bar z)\equiv H_\psi(q,p)=|\braket{z}{\psi}_d|^2\rme^{-\pi d z\bar z} \,,
\ee
is positive and strictly periodic because of (\ref{eq:boundary}). The important observation \cite{Leboeuf_1990} is that for pure states, 
this distribution inherits the zeroes of the Bargmann function and thus vanishes exactly at $d$ points. It factorizes into products of
elementary Husimi functions 
\be
H_\psi(z,\bar z)=\prod_{i=1}^d \rme^{-\pi z\bar z}\left|\theta_3\big(-\rmi\pi (z+z_0-z_i)|\rmi \big)\right|^2 \,.
\ee
In contrast with the Wigner and chord distributions, the pure state condition for the Husimi function is easily represented by this 
factorization formula, which does not hold for mixed states. Thus, a given pure state is uniquely defined by the distribution of $d$ 
zeroes in the unit cell, constrained by having their center of mass in the center of the cell, Eq. (\ref{eq:constraint}).  
Understanding the special relations among the positions of these zeroes that make $\ket{\psi}$ a fiducial state is another way of stating the difficulty of the SIC-POVM problem.


As an illustration, Fig. \ref{husbarg} shows the Bargmann and Husimi representations of the state ``5a'' in the list in \cite{Scott_2010} 
(the same state whose Wigner and chord representations are given in Fig. \ref{chordwigner}). The zeroes of the Husimi density (bottom panel) 
can be identified as dislocations in the phase of the Bargmann function (top panel). 

Husimi, chord and Wigner representation for odd-dimensional states up to $d=67$ are available online at \cite{Oddrepres}.

\section {Localization measures} \label{sec: localization}

In this section we use the fact that a SIC-POVM set is a two-design \cite{Klappenecker_2005} to study the localization properties of 
fiducial states. We take as a measure of localization the inverse participation ratio, which gives an idea of how ``concentrated'' a 
normalized probability distribution $p_i,\, i=1\cdots K$ is in its probability space. It is defined as:
\be
P=\sum_{i=1}^K p_i^2\,.
\ee
For a normalized pure state $\ket{\psi}\in {\cal H}_d$ it takes the form
\be
\label{eq: IPR}
   P_\psi=\sum_{i=0}^{d-1} |\braket{i}{\psi}|^4
\ee
Its value ranges from $P=d^{-1}$ for a state uniformly spread out in
the basis considered, with 
$|\braket{i}{\psi}|=1/\sqrt{d}~\forall~i$, to $P=1$ for each of the basis
states. This quantity is strongly dependent on the 
chosen basis and in general has no invariant significance under unitary transformations
of the state. 

The inverse participation ratio for a fiducial state $\ket{\phi}$ follows from the two-design property
\ba
\frac{1}{d^2}\sum_{\bfalpha\in {\mathbb Z}_{d}^2}&&\bra{\phi}\optr{\bfalpha}^\dagger\hat A_1\optr{\bfalpha}\ket{\phi}\bra{\phi}\optr{\bfalpha}^\dagger\hat A_2\optr{\bfalpha}\ket{\phi}\nonumber\\
&=& \int\rmd\psi \matel{\psi}{A_1}{\psi} \matel{\psi}{A_2}{\psi}
\label{eq:two-design}
\ea
valid for arbitrary $\hat A_1, \hat A_2$ . The integral on the r.h.s. is over the normalized Haar measure and can be evaluated as 
\cite{Emerson_2005, Dankert_2009}:
\be
 \int\rmd\psi \matel{\psi}{A_1}{\psi} \matel{\psi}{A_2}{\psi}=\frac{\tr\big(\hat A_1\hat A_2\big)+\tr\big(\hat A_1\big)\tr\big(\hat A_2\big)}{d(d+1)}
 \label{eq:two-design-result}
\ee
Replacing $\hat A_1 = \hat A_2=\ket{i}\bra{i}$ with $\ket{i}$ any of the basis states and combining (\ref{eq:two-design}) and (\ref{eq:two-design-result}) one obtains:
\be
\frac{1}{d^2}\sum_{\bfalpha\in {\mathbb Z}_{d}^2}|\bra{i}\optr{\bfalpha}\ket{\phi}|^4 = \frac{2}{d(d+1)}
\ee
which directly shows that:
\be
P_{\rm SICPOVM}= \sum_{i=0}^{d-1} |\braket{i}{\phi}|^4=\frac{2}{d+1}\,.
\ee
This result is the same in all other bases related by Clifford operations.

Making the same replacement in (\ref{eq:two-design-result}) one also sees that 
\be
\int\rmd\psi |\braket{i}{\psi}|^4 =\frac{2}{d(d+1)},
\ee
which also implies:
\be
<P>_{Haar}=\int\rmd\psi \sum_{i=0}^{d-1} |\braket{i}{\psi}|^4 =\frac{2}{d+1}.
\ee
This means that the average value of the inverse participation ratio over random states distributed over the Haar measure is the 
same as its value for SIC-POVM fiducial states, with a localization half way between maximum and minimum. This is illustrated in 
Fig. \ref{randomposchord} (top), which shows the degree of delocalization for different kinds of pure quantum states; for the 
sake of clarity we plot $P^{-1}$, a quantity that ranges between 1 and $d$, instead of $P$. 

\begin{figure}[h]
\begin{center}
\includegraphics*[width=0.45\textwidth]{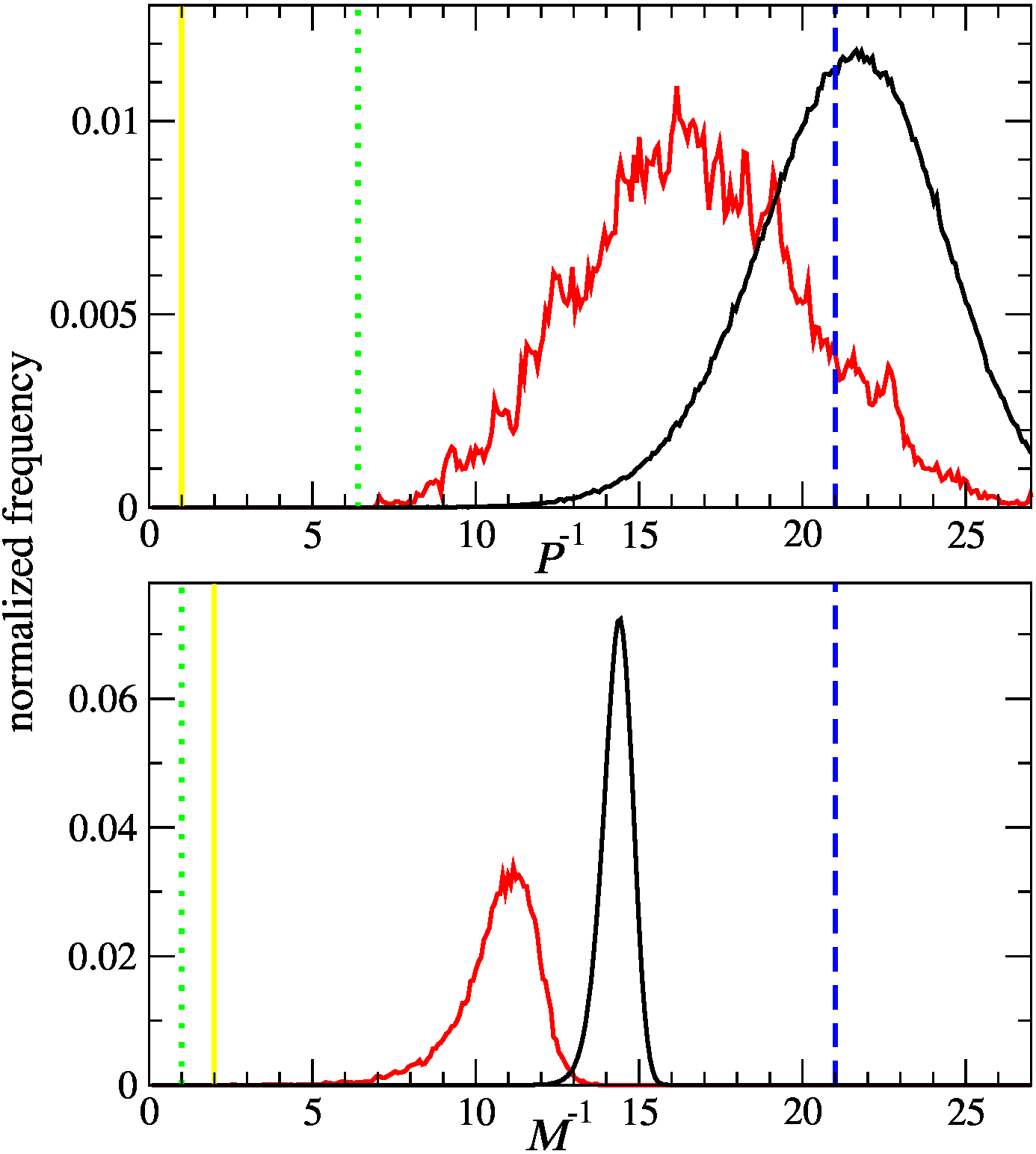}
\caption{(color online) Measures of delocalization for different quantum states for a Hilbert space of dimension $d=41$.
The top shows $P^{-1}$, with $P$ the inverse participation ratio defined in Eq.~(\ref{eq: IPR}), and the bottom panel the phase-space 
delocalization measure $M^{-1}$, with $M$ from Eq.~(\ref{eq: spread}). The solid lines show the histograms for $10^6$ random pure states 
(black), and eigenstates of a chaotic 
Harper map (red) with 500 bins in $P^{-1},M\in[0,41]$. 
Position, coherent and SIC-POVM fiducial states are represented in solid yellow, dotted green and dashed blue vertical lines respectively.   
\label{randomposchord}}
\end{center}
\end{figure}

We now want to extend these considerations to phase-space localization. For the study of SIC-POVM fiducial states, we would like to define
localization measures that share the invariance under
Clifford operations. To this end we consider first the
chord distribution, as defined in Eqs. (\ref{eq: chord}-\ref{eq: state chord}). 
For a pure state, the condition \mbox{$\tr (\rho^2)=1$} leads to the
normalization $1/d \, \sum_\bfalpha |C(\alpha)|^2=1 $. A suitable phase-space measure for localization 
can then take the form:
\be
M_\psi =\frac{1}{d}\sum_{\bfalpha\in {\mathbb Z}_{d}^2} |C_{\psi}(\alpha)|^4=\frac{1}{d}\sum_{\bfalpha\in {\mathbb Z}_{d}^2} |\bra{\psi}\optr{\bfalpha}\ket{\psi}|^4 \,.
\label{eq: spread}
\ee
In the same vein we can also consider the Wigner localization as:
\be
M_\psi =\frac{1}{d}\sum_{\bfalpha\in {\mathbb Z}_{d}^2} W_{\psi}^4(\mbfx)=\frac{1}{d}\sum_{\bfalpha\in {\mathbb Z}_{d}^2} \bra{\psi}\opref{\mbfx}\ket{\psi}^4 \,.
\ee
In Appendix \ref{sec:equivalence} we show that, for pure states, these two definitions indeed coincide; besides, the covariance of 
$ \optr {\bfalpha}$ and $\opref{\mbfx}$ under Clifford operations shows that $M$ is invariant under them as desired. $M$ satisfies the bounds 
\be
           \frac{2}{d+1} \le M \le 1 \,.
\ee           
The upper bound is a simple consequence of the fact that $|\matel{\psi}{T_\bfalpha}{\psi}|\le 1$ and is saturated, i.e, by position 
states $\ket{\psi}=\ket{i}$.
The lower bound is more subtle \cite {Welch} and the important result is that it is attained if and only if $\ket{\psi}$ is a fiducial 
state \cite{Renes_2004}. In fact $M$, written in a slightly different fashion, is the cost function used for numerical searches \cite{Scott_2010}.
Thus, with respect to this measure of phase-space localization, we can identify SIC-POVM fiducial states as those maximally delocalized in phase space.

The Haar average of $M$ can also be computed with the techniques of \cite{Brouwer} with the result:
\be
<M>_{Haar}=\int \rmd\psi M_\psi =\frac{3}{d+2}
\label{eq:mhaaraverage}
\ee 
(more details are given in Appendix \ref{sec:averages}). So the phase-space localization properties of random states are clearly different
from those of SIC-POVM fiducial states, in contrast to the behavior observed for localization in a given basis of the Hilbert space. 
The phase-space localization properties of different states are compared in Fig.~\ref{randomposchord} (bottom panel), where for clarity we plot $M^{-1}$ instead of $M$.


\section{The Zauner symmetry} \label{sec: Zauner}

\subsection{Zauner operator and its classical counterpart}

In his Ph.D. thesis \cite{Zauner_2011}, Zauner conjectured that the fiducial
state of a Heisenberg-Weyl 
SIC-POVM was to be found as a particular eigenfunction of
a unitary map. 
In the position representation, the map has the following
matrix elements
\be
\matel{k}{Z}{j}=\frac{\rme^{\rmi\chi}}{\sqrt d} \, \tau^{2kj
+k^2}
\ee
Choosing the phase $\chi=\pi(d-1)/12$ the map has the
property $\hat Z^3=1$. 
Its eigenvalues are thus $\cep{2\pi n/3}~ (n=0,1,2)$ and
the spectrum is highly degenerate for all large values of $d$. 
Because of the large degeneracy, the Zauner
hypothesis puts a restriction 
on the fiducial state but by no means determines it
completely. Finding a fiducial state can thus be considered as
the numerical task of finding the particular linear
combination of eigenstates of the Zauner map within one of its three multiplets and satisfying Eq.\,(\ref{eq: fiducial state}).
The map has two symmetries: a unitary symmetry under parity and an antiunitary one under time reversal. These are reflected in the properties
\be
\opref{0}\hat Z \opref{0}= \hat Z, ~~~~~~~~ \hat F \hat Z \hat F^\dagger =( \hat Z^\ast)^{-1} 
\ee
where the asterisk denotes complex conjugation and $\hat F$ is the Fourier operator $\matel{i}{F}{j}=\omega^{-ij}/\sqrt{d}$.

The Zauner operator can be considered as the quantum version of a classical 
map which is a symplectic 
 automorphism of a 2D torus. The map is defined on the square 
 $E = [-\half,\half)\times [-\half,\half)$ and maps $(q,p)\in E$ to $(q',p')\in E$
 according to:
\ba
   q'&=&-p\\  
   p'&=& q-p -\floor{2(q-p)}  
   \label{zaunermap}
\ea
where $\floor{\cdot}$ denotes the integer part. It is an area-preserving 
elliptic ``cat map'' which is locally a composition of
a negative shear followed by a $\pi/2$ rotation as
illustrated in Fig.\,\ref{zauner}.

\begin{figure}
\begin{center}
\includegraphics[width=0.45\textwidth]{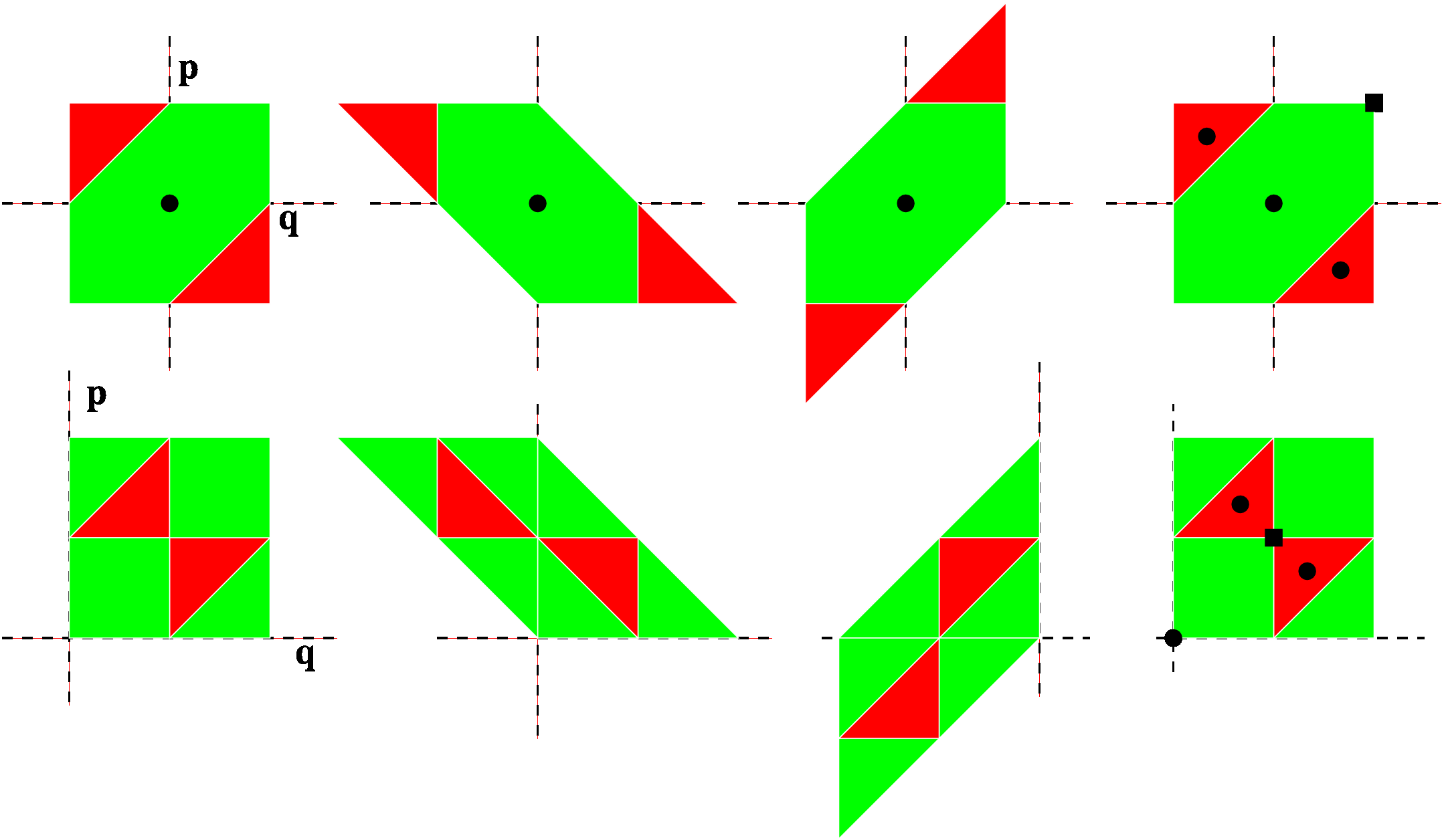}
\caption{
Classical Zauner map representation in phase space. 
Invariants regions are shown in colors with
fixed points represented by black dots.
The top and bottom rows represent the action of the
map for two equivalent choices of the phase space 
where origin is on the center and left bottom respectively.
\label{zauner}}
\end{center}
\end{figure}

The map has three invariant regions $E_0,E_1,E_{-1}$
which are determined by the value of $\epsilon=\floor{2(p-q)}
=0,1,-1$. These regions are shown in Fig.\,\ref{zauner}; for clarity, in the bottom row we show the map acting on the square $[0,1)\times [0,1)$. 
All orbits are of period three and their
respective cycles are
\be
\vect{q}{p}\to\vect{-p}{q-p-\epsilon}
\to\vect{p-q+\epsilon}{-q}\to\vect{q}{p}
\label{cycle}
\ee
There are three fixed points belonging to
each of the invariant regions $E_i$: 
$z_0=(0,0)$, $z_1=(1/3,-1/3)$, $z_{-1}=(-1/3,1/3)$. Moreover, the point $(q,p)=(\half,\half)$  can be assimilated to a hyperbolic point 
with $q=\half$ and $p=\half$ as stable and unstable manifolds. The respective areas of the invariant regions are $\mu(E_1)=\mu(E_{-1})=1/8$ 
and $\mu_0=3/4$, with a total unit area. 

The quantum Zauner map is a special Clifford operation that shares the basic property of mapping the Weyl operators among themselves, i.e.
\be
\hat Z \optr{\bfalpha}\hat Z^\dagger= \optr{Z\bfalpha}
\ee
where $Z$ is the discretized map acting as
\be
Z\bfalpha=\mat{0}{-1}{1}{-1}\vect{\alpha_1}{\alpha_2}  \mod d\,.
\ee
The correspondence with the classical map is obtained discretizing $ q,p \to \alpha_1/d,\alpha_2/d$ with
\be
-(d-1)/2\le\alpha_1,\alpha_2\le (d-1)/2. 
\ee

The invariant regions are preserved by the discretization and are now defined as $\bar E_\epsilon$ where $\epsilon=\floor{2(\alpha_1 -\alpha_2)/d}$.
The totality of $d^2$ points are divided in three invariant sets of dimension 
\ba
\mu(\bar E_1)&=&\mu(\bar E_{-1})= \frac{(d-1)(d+1)}{8}\nonumber\\
\mu(\bar E_0)&=& d^2 - 2\frac{d^2-1}{8} = \frac{3d^2+1}{4}
\ea
in close correspondence with the classical areas, if we assign to each point an area $d^{-2}$.
We display the discrete invariant regions in Fig.\,\ref{discretezauner}.
\begin{figure}
\begin{center}
\includegraphics[width=0.4\textwidth]{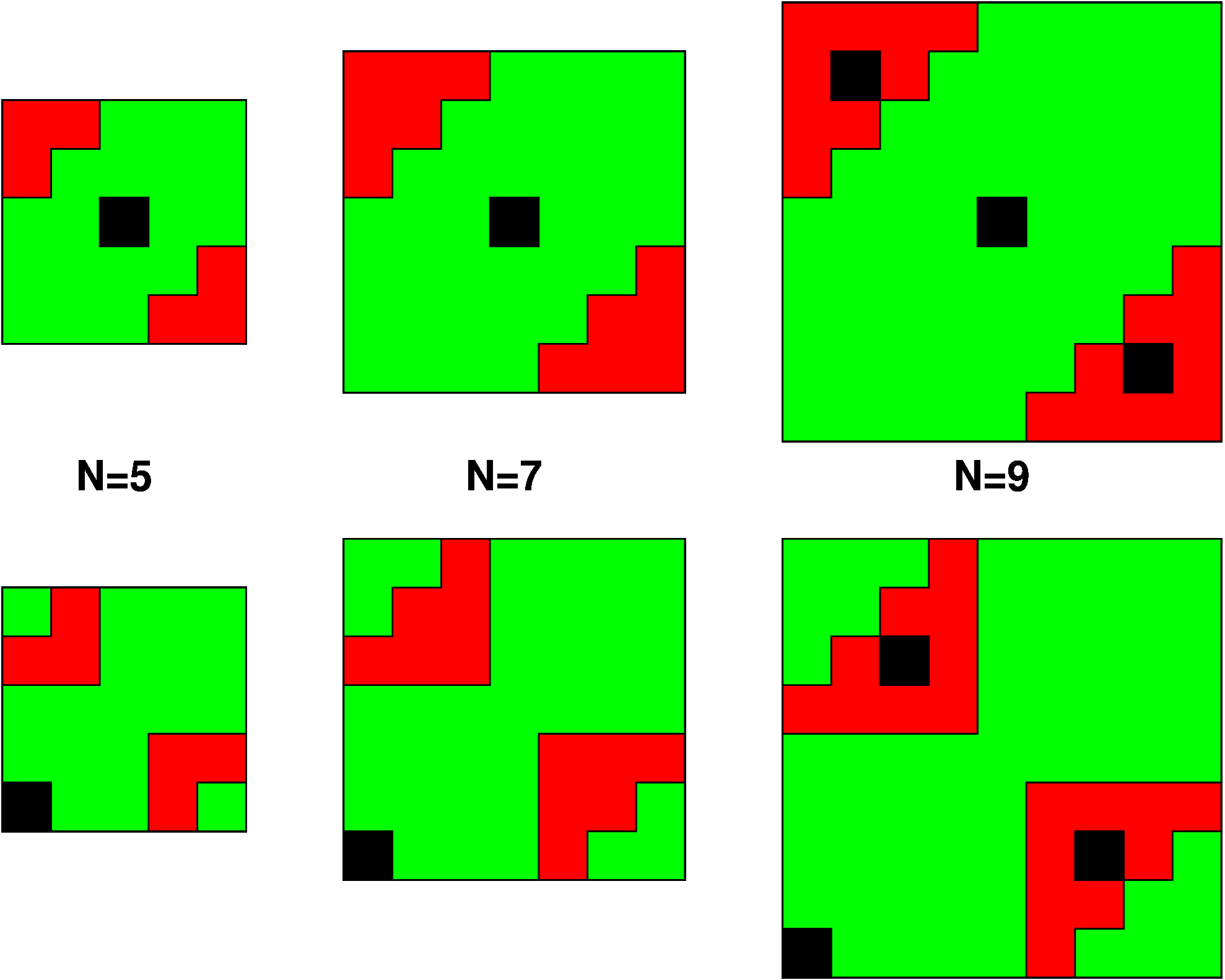}
\caption{ 
Invariants regions for the discretized classical Zauner map. There are three fixed points if $d$ is divisible by 3, 
and only one otherwise. As in the previous Figure, top and bottom rows correspond to two equivalent choices for the phase-space limits
where origin is on the center and left bottom respectively.
\label{discretezauner}}
\end{center}
\end{figure}
The fixed point at $\bfalpha=(0,0)$ in the central region $\bar E_0$ survives the discretization for all odd $d$, while the the other two fixed 
points only survive when $d$ is an odd multiple of $3$, and they occur at $\bfalpha_1=(d/3,-d/3)$ and $\bfalpha_{-1}=(-d/3,d/3)$. 

Taking into account that $Z^3=1$ we can divide each region into 3-cycles (and fixed points) as follows. It is easy to calculate the number $N_i$ 
of 3-cycles in each region. In $\bar E_0$ (excluding the trivial fixed point) the count is as follows:
\be
N_0=\frac{1}{3}\left( \frac{3d^2+1}{4}-1\right)=\frac{d^2-1}{4}
\label{eq:cyclecount1}
\ee
These cycles are further divided in two disjoint sets related by the symmetry $\bfalpha\to -\bfalpha$.
In regions $\bar E_{\pm1}$ the count depends on whether 3 divides $d$
\be
N_{\pm1}=\left\{
\begin{array}{ll}
\frac{1}{3}\left(\frac{d^2-1}{8}-1\right)=\frac{d^2-9}{24} \phantom{\quad} &{\rm for}~ d/3~{\rm integer} \vspace{3pt} \\
\frac{d^2-1}{24} &{\rm otherwise} 
\end {array}
\right.\label{eq:cyclecount2}
\ee

These considerations are relevant for the structure of a fiducial state in the chord representation. In fact, the Zauner symmetry 
$\hat Z\hat\sigma\hat Z^\dagger=\hat\sigma$ imposed on Eq.\,(\ref{chordstructure}) implies that the phases $\psi_\bfalpha$ must be constant 
along each of the 3-cycles. Then $\hat\sigma$ is further constrained as
\be
\hat\sigma=\frac{1}{d}
\left[ 
\hat 1 +\frac{1}{\sqrt{d+1}}\sum_p 
\frac{\rme^{\rmi\psi_p}\hat M_p + \rme^{-\rmi\psi_p} \hat M_p^\dagger}{2}
\right]
\label{eq:cyclestructure}
\ee
where $p$ runs over all cycles other than the fixed point at the origin. In this expression we have explicitly used the phase relation between 
cycles reflected through the origin, and defined:
\be
\hat M_p= \sum_{\bfalpha\in p} \optr{\bfalpha}\,.
\ee
Notice that each cycle can correspond to three points or only one, if $d$ is divisible by 3. The total number of phases $\# p$ not constrained 
by the Zauner symmetry can be computed from Eqs.~(\ref{eq:cyclecount1}-\ref{eq:cyclecount2}) and yield 
\be
\# p=\left\{
\begin{array}{ll} 
                  \frac{d^2+3}{6}  \phantom{\quad} &{\rm for}~ d/3~{\rm integer} \vspace{3pt} \\
                  \frac{d^2-1}{6}  & {\rm otherwise}
\end{array}
\right.                  
\ee
The analysis of the Wigner case proceeds in a similar way. Again the fact that  $\hat Z \opref{\mbfx}\hat Z^\dagger=\opref{Z\mbfx}$ implies that 
the values of the Wigner function are constant on the cycles, although here there is no direct relationship between $W(\mbfx)$ and $W(-\mbfx)$. 
All there features are clearly seen in Fig.\ref{chordwigner}, where we have marked the cycles joining equal values of the distributions.

\subsection{Definition of a basis of eigenvectors of the Zauner map}

To specify completely an eigenbasis of $\hat Z$ one needs to 
find another operator commuting with it that can remove the 
degeneracies. We have constructed such an operator 
as a superposition of translations on one of the classical 
periodic orbits of $ Z$:
\be
\hat H^\bfalpha=\frac{1}{2}\left(\optr{\bfalpha}+\optr{Z\bfalpha}
+\optr{Z^2\bfalpha}+{\rm H.c.}\right)\,.
\ee
The operator is labeled by one point on the orbit and is clearly Hermitian. Obviously $\hat Z H^\bfalpha \hat Z^\dagger =H^\bfalpha$. Moreover 
from the fact that $\opref{0}\optr{\bfalpha}\opref{0}=\optr{-\bfalpha}$, $\hat H^\bfalpha$ also commutes with the parity operation. Thus the 
three commuting operators $\hat Z,\hat H^\bfalpha, \opref{0} $ have common eigenfunctions defined by the properties:
\ba
\hat H^\bfalpha& \ket{\psi_{ikr}}&=\epsilon_{ikr}\ket{\psi_{ikr}}\\
\opref{0}& \ket{\psi_{ikr}}&=r\ket{\psi_{ikr}}\\
\hat Z &\ket{\psi_{ikr}}&=\rme^{2\pi\rmi k/3}\ket{\psi_{ikr}}
\ea
where $r=\pm 1$, $k=0,1,2$ and $\epsilon_{ikr}$ are the non-degenerate eigenvalues of $\hat H^\bfalpha$.

We now take the simplest choice of such Hamiltonian, built on the orbit $(0,1),(1,0),(-1,-1)$ leading  to
\be
\hat H= \frac{\hat U +\hat U^\dagger}{2} 
+ \frac{\hat V +\hat V^\dagger}{2} + \frac{\hat U\hat V \tau^\ast+\hat 
V^\dagger\hat U^\dagger\tau}{2} \,.
\label{eq:qHarperHam}
\ee
The resulting hermitian matrix is easily constructed and can be numerically diagonalized. In the limit of $d\to \infty$ it corresponds to the
Weyl quantization of the classical Hamiltonian  
\be
H(q,p)=\cos(2\pi q)+\cos(2\pi p)-\cos[2\pi (q-p)].
\label{hamiltonian}
\ee
where $q,p\in [0,1)$ are now continuous variables corresponding to $\alpha_1/d,\alpha_2/d$. This is a variant of the Harper Hamiltonian on the 
torus, in both its classical an quantum versions. In Fig.\,\ref{zaunerinvariant} we plot the level curves of $H(q,p)$, a smooth doubly periodic 
surface which displays very similar features as the map (cf. Fig.\,\ref{discretezauner}), sharing the invariant regions and the fixed points as 
critical points of the surface. The level curves $H(q,p)=E$ support semiclassically the eigenstates of the quantum Hamiltonian, with eigenvalues  
approximately quantized by the Bohr-Sommerfeld rule
\be
S(E)=\oint_E p(E,q)\rmd q = \frac{2\pi}{d}\left(\rmi+\half\right)\,.
\ee
It is easy to check that the three points of each cycle of the classical map lie on one of these levels and each level supports a continuous 
family of such cycles. 

\begin{figure}[hbt]
\begin{center}
\includegraphics[width=0.35\textwidth]{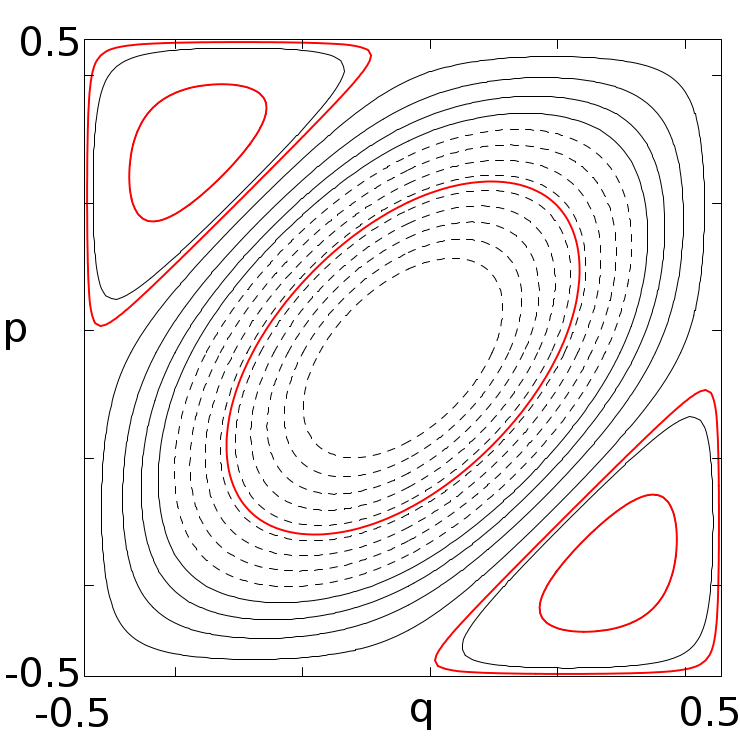}
\caption{%
(color online)Level curves for the classical Hamiltonian  defined in Eq. (\ref{hamiltonian}). 
 In boldface (red online) we display the three levels with energies that support the $k=0$ symmetry for the state ``7a''.
\label{zaunerinvariant}}
\end{center}
\end{figure}

In table \ref{tab:eigenvalues} we show the numerically determined eigenvalues and quantum numbers of $\hat H$ for $d=7$. As an example, 
the fiducial state ``7a'' \cite{Scott_2010}, which belongs to the multiplet $k=0$, has the expansion
\be
\ket{7a}=a_1\ket{\psi_{101}}+a_5\ket{\psi_{50-1}}+a_6\ket{\psi_{601}}
\ee
where $a_1\simeq0.336$, $a_5\simeq-0.691+\rmi \,0.230$, and $a_6\simeq -0.107+\rmi \,0.587$. Thus, contrary to the representation in (\ref{eq:cyclestructure}) in terms of classical cycles of the map, here the fiducial states are 
represented as superpositions of eigenfunctions of the quantum Hamiltonian, which have a clear semiclassical ($d\to\infty$) description 
in terms of ``tori'' of the classical Hamiltonian.

\begin{table}
\begin {tabular}{|c|c|c|c|}
\hline 
$i$   &  $\epsilon_i$  & $ r$     & $k$    \\
\hline
0   & -2.315069600541   & -1 &-1    \\
 1  &  -1.118527682059  &   1 & 0    \\
 2  &  -0.209389069220  &  -1 & 1    \\
  3   &  0.337407948091 &    1 &-1  \\  
 4    & 0.940071117953   &  1 & 1    \\
 5   &  1.024458669761  &  -1 & 0  \\  
 6    & 1.341048616015   & 1  &0  \\
 \hline   
  \end{tabular}
\caption{Eigenvalues of Hamiltonian $\hat H$ from Eq.\,(\ref{eq:qHarperHam}), and their corresponding quantum numbers $r$ and $k$ under 
parity and Zauner map respectively, for $d=7$.}
\label{tab:eigenvalues}
\end{table}



\section{Concluding remarks} \label{sec: conclusions}

We have explored several ways of representing the fiducial states using phase-space methods. The discrete representations based on the 
chord and Wigner functions are very similar and (for odd $d$) yield $d\times d$ arrays respectively of phases with constant amplitudes or 
real (positive and negative) values. Clifford symmetries of the fiducial state thus represented are readily recognized because the array 
values are constant on the cycles of the classical symmetry. In particular, the presence of the Zauner symmetry organizes the distributions 
as superpositions of 3-cycles as in Eq.\,(\ref{eq:cyclestructure}). However, the representation is highly redundant, as it depends on a 
number of parameters growing as $d^2$, even taking into account the presence of symmetries, while the state itself only depends on $d-1$
complex amplitudes. The imposition of the nonlinear pure-state condition directly in terms of these two quasidistributions remains elusive.

The continuum representation is much leaner, and in a way has complementary properties. The pure-state condition is readily accommodated 
by a constellation of \mbox{$d-1$} independent zeroes that parametrize the fiducial state in a way equivalent to that provided by the 
representation in terms of amplitudes. However we have not found a clear signature in this constellation of the presence of a Clifford 
symmetry, besides of course the obvious invariance of the whole constellation. The reason is that, unlike what happens for the discrete 
representations, the zeroes do not ``move'' classically when the state is subjected to a Clifford operation. It remains also a major 
open issue to precisely formulate the fiducial SIC-POVM condition directly in terms of the zeroes of the constellation.

The fact that a SIC-POVM is a projective 2-design imposes some constraints on the localization properties of the fiducial states, as 
measured by the inverse participation ratio. When the fiducial state is expanded in the coordinate basis, its inverse participation ratio 
$P$ has the same value as the average over random states. On the other hand, when fiducial states are described by means of the chord or the Wigner representations, the inverse participation ratio shows they are maximally delocalized in phase space, in contrast with the Haar average. 

The conjectured Zauner symmetry shows up in interesting ways in the discrete representations, its main consequence being that the 
distributions become constant along the classical $3$-cycles of the symmetry. The fiducial state is then represented as a superposition 
of the order of $(d^2-1)/3$ periodic cycles. In contrast, the representation as a superposition of the eigenstates of a Hamiltonian 
$\hat H$ commuting with the Zauner operator, has the least possible freedom compatible with the symmetry ($\approx d/3$). Moreover 
the eigenstates retain  simple semiclassical features that become sharper as $d\to\infty$.

In summary,  we have explored various ways to display phase-space features of SIC-POVM fiducial states. Although none provides significant 
inroads to the deeper problem of the proof of existence,  we present them in the hope that this study may illuminate from different angles 
some of the many facets of this most interesting problem.

\appendix

\section{Equivalence for pure states of the two definitions of the phase-space localization measure}
\label{sec:equivalence}

To show that the two definitions of $M $ in Eq.~(\ref{eq: spread}) are indeed equal we consider, for an arbitrary operator $\hat A$, the quantity
\be
\tr(\opref{\mbfx}\hat A \opref{\mbfx} \hat A)=\frac{1}{d^2}\sum_{\bfalpha,\bfbeta}~C_A(\bfalpha)C_A(\bfbeta)\tr(\opref{\mbfx}\optr{\bfalpha}\opref{\mbfx}\optr{\bfbeta})
\ee
where we have expanded $\hat A$ in terms of the chord function using Eq. (\ref{eq: state chord}). Using the conjugation
\be
\opref{\mbfx}\optr{\bfalpha}\opref{\mbfx}=\omega^{2\sym{\bfalpha}{\mbfx}}\optr{\bfalpha}^\dagger
\ee
and the orthogonality of translations we deduce
\be
\tr(\opref{\mbfx}\hat A \opref{\mbfx} \hat A)=\frac{1}{d}\sum_{\bfalpha}~C_A(\bfalpha)C_A(\bfalpha)\omega^{2\sym{\bfalpha}{\mbfx}}
\ee
Now, if $\hat A$ is a pure state $\ket{\psi}\bra{\psi}$ the l.h.s. is the square of the Wigner function $W_\psi(\mbfx)=\matel{\psi}{R_{\mbfx}}{\psi}$ and we have obtained
\be
W_\psi(\mbfx)^2= \frac{1}{d}\sum_{\bfalpha}\omega^{2\sym{\bfalpha}{\mbfx}} C_\psi(\bfalpha)^2
\ee
Thus the square of the Wigner function is the symplectic Fourier transform of the square of the chord function. Parseval relation then immediately implies
\be
\sum_{\mbfx}W_\psi(\mbfx)^4=\sum_{\bfalpha} |C_\psi(\bfalpha)|^4
\ee

\section{Averages over Haar measure}
\label{sec:averages}

We are interested in averages of the type
\be 
  \int\diff\psi~\prod_{i=1}^t\matel{\psi}{ A_i}{\psi}
\ee
where the integration is over the complex projective space $CP^{d-1}$ with the measure induced by the unitarily invariant Haar measure on $U(d)$.
The symmetry of the integrand under the interchange of the $\hat A_i$ turns this into a projector onto the totally symmetric 
subspace of the Hilbert space ${\cal H}^{\otimes t}\equiv C_d^{\otimes t}$ with dimension 
\be 
D_t=\frac{(d+t-1)!}{(d-1)!~t!}
\ee

The average can be expressed as a trace over permanents
\begin{widetext}
\be 
 \int\diff\psi~\prod_{i=1}^t\matel{\psi}{ A_i}{\psi}=
\frac{(d-1)!}{(d+t-1)!}\sum_{i_1i_2\cdots i_t}\left|
\begin{array}{cccc}
 \matel{i_1}{A_1}{i_1} &\matel{i_1}{A_2}{i_2} &\cdots &\matel{i_1}{A_t}{i_t} \\
 \matel{i_2}{A_1}{i_1} &\matel{i_2}{A_2}{i_2} &\cdots &\matel{i_2}{A_t}{i_t} \\
 \vdots                &  \vdots               & \ddots& \vdots                \\
 \matel{i_t}{A_1}{i_1} &\matel{i_t}{A_2}{i_2} &\cdots &\matel{i_t}{A_t}{i_t}
\end{array}
\right|_+ \label{eq.average}
\ee
\end{widetext}

The permanent can be evaluated in terms of all possible product of traces of the $A_i$. For $t=2$ and for traceless and unitary translations 
with $A_1=A_2=\optr{\bfalpha}, A_3=A_4=\optr{\bfalpha}^\dagger$, Eq.\,(\ref{eq.average}) can be easily evaluated with the result:
\be
 \int\diff\psi~\left|\matel{\psi}{T_\bfalpha }{\psi}\right|^4= \left\{
\begin{array}{cr}
\frac{2}{(d+1)(d+2)} &{\rm for}~~ \bfalpha\ne0 \\
 1              & {\rm for}~~ \bfalpha=0
\end{array}
\right.
\label{eq: epsilon2}
\ee

Then we obtain Eq.\,(\ref{eq:mhaaraverage})
\be
<M>_{Haar}=\frac{1}{d}\left\{1+\frac{2(d^2-1)}{(d+1)(d+2)}\right\}= \frac{3}{d+2}
\ee


\end{document}